\documentclass[aps,nofootinbib,showpacs,preprintnumbers,amsmath,amssymb]{revtex4}
\def\be{\begin{equation}}
\def\ee{\end{equation}}
\def\ba{\begin{eqnarray}}
\def\ea{\end{eqnarray}}
\def \bea{\begin{eqnarray}}
\def \eea{\end{eqnarray}}

\usepackage{graphics}
\usepackage{graphicx}% Include figure files
\usepackage{dcolumn}% Align table columns on decimal point
\usepackage{bm}% bold math
\usepackage{epsfig}
\usepackage{graphicx}
\usepackage{multirow}
\usepackage{dcolumn}% Align table columns on decimal point
\usepackage{graphicx,epsfig}%
\def \ee{\end{equation}}
\def \be{\begin{equation}}
\def \bea{\begin{eqnarray}}
\def \eea{\end{eqnarray}}

\preprint{}
\begin{document}

\title{Exact black hole solution for scale dependent gravitational couplings
and the corresponding coupling flow}

\author{Carlos Contreras$^\dagger$, Benjamin Koch$^*$,  Paola Rioseco$^*$}
 \affiliation{
$^\dagger$ Departamento de F\'{i}sica, Universidad T\'{e}cnica Federico Santa Mar\'{i}a;\\
Casilla 110-V, Valpara\'{i}so, Chile;\\
$^*$ Pontificia Universidad Cat\'{o}lica de Chile, \\
Av. Vicu\~{n}a Mackenna 4860, \\
Santiago, Chile \\
}
\date{\today}

\begin{abstract}
We study a black hole solution for the generalized Einstein Hilbert
action with scale dependent couplings $G(r)$ and $\Lambda(r)$.
The form of the couplings is not imposed, but rather deduced
from the existence of a non trivial symmetrical solution.
A classical-like choice of the integration constants is found.
Finally, the induced flow of the couplings is derived and compared
to the flow that is obtained in the context of the exact renormalization group
approach.
\end{abstract}

\maketitle

%%%%%%%%%% FIGURE %%%%%%%%%%%%%%%%%%

%%%%%%%%%%%%%%%%%%%%%%%%%%%%%%%%%%%%%%%%%%%%

%
%%%%%%%%%%%%%%%%%%%%%%%%%%%%%%%55
\section{Introduction}
One of the many achievements of quantum field theories like the standard model
is the confirmed prediction of a scale dependence of the physical couplings $\alpha \rightarrow \alpha_k$.
In order to know in which way the couplings $\alpha$ of a given quantum field theory depend on the 
energy scale $k$ one usually has to regularize and renormalize the theory.
However, when it comes to gravity a consistent and predictive renormalization and therefore
a predictive quantum field theoretical description is still to be found.
Independently of how this theory of quantum gravity will look like, 
in most approaches it
is expected to introduce a non-trivial running to the couplings of classical gravity
which are Newtons ``constant'' $G\rightarrow G(k)$ and the cosmological
constant $\Lambda \rightarrow \Lambda(k)$. In an effective description
those couplings are expected to be present in an improved action and the corresponding
solutions.

Since on the one hand
black holes are key objects in every classical or quantum gravitational theory
and their understanding is crucial for the understanding of the whole model
and on the other hand
the scale dependence of gravitational couplings has been extensively studied
in the context of exact renormalization group (ERG) equations 
\cite{Weinberg:1979,Wetterich:1992yh,Dou:1997fg,Souma:1999at,Reuter:2001ag,Litim:2003vp,Fischer:2006fz,Percacci:2007sz,Litim:2008tt,Narain:2009gb,Groh:2010ta,Bonanno:2001hi,Reuter:2012id,Reuter:2012xf,Benedetti:2013jk},
it is natural to study black holes in the context of ERG results.
This has been done mostly by improving the classical solutions
\cite{Bonanno:1998ye,Bonanno:2000ep,Reuter:2003ca,Emoto:2006vx,Girelli:2006sc,Codello:2007bd,Koch:2007yt,Koch:2008zzb,Burschil:2009va,Falls:2010he,Casadio:2010fw,Reuter:2010xb}.
This procedure has however two weaknesses which basically motivated this study:
The first problem is that in order to study black holes in the ERG context one has to relate the radial scale of a black hole solution $r$ to the energy scale $k$ of the ERG calculation, this procedure is however not uniquely defined.
The second problem is that the improved solution does not
(at least not at all scales \cite{FrankBen}) resolve the improved
equations of motion nor does it minimize the ERG improved action.

As complementary contribution to this program we will follow a philosophy that
is somewhat inverse to the existing studies on ERG improved black hole solutions.
As starting point we will take the improved equations of motions which contain
scale depending couplings $G(r)$ and $\Lambda(r)$ which are a priory undetermined.
The working hypothesis will then be to ask for which functional form of those
couplings it is possible to solve those equations of motion with the most
symmetrically possible metrical ansatz.

The paper is organized as follows.
In section \ref{SecSol} it is shown how solving the resulting system of equations
determines then the resulting black hole metric and the functional form
of $G(r)$ and $\Lambda(r)$ up to the existence of four integration constants.
General properties of this solution such as differences to the studies in the literature,
singularities, and the existence of classical-like parameter choices
 are then discussed.

Since both couplings of the present solution are functions of the radial scale $r$,
the corresponding adimensional couplings $g(r)$ and $\lambda(r)$ can be
combined in a coupling flow. In section \ref{SecRun} this induced flow with an
ultra violet fixed point for
 $g(r)$ and $\lambda(r)$ is derived, the flow is compared to the flow
 for $g_k$ and $\lambda_k$ known from ERG calculations.
Finally, the anomalous dimensions of the induced couplings
and the product of couplings are discussed and
and compared to the findings in the ERG approach.

After summarizing remarks in section \ref{SecConcl}
we give complementary discussions in the appendix \ref{SecAppend}.

\pagebreak
%%%%%%%%%%%%%%%%%%%%%%%%%%%%%%%%%%%%%%%%%5
\section{Exact solution with cosmological term}\label{SecSol}
\subsection{A black  hole solution}
Up to now black holes where studied in the context of ERG equations by
taking the classical solutions (for constant
$G$ and $\Lambda$) and then replacing $G\rightarrow G(k)$
and $\Lambda\rightarrow \Lambda(k)$ in those solutions.
The physical interpretation of those ERG improved solutions
depends on how the ERG scale $k$ is related to the physical scale $r$.
This procedure however was only partially successful, since
the improved solutions were actually no more
solutions of any form of Einstein or modified Einstein equations.
Yet, the improved solution with a cosmological term seems at least
to solve the improved equations of motion asymptotically in the UV \cite{FrankBen}.

In order to avoid the danger which is involved with the choice
of $k\rightarrow k(r)$ we will consider directly the fact
that all scales are in the end functions of the physical scale
$G=G(r)$ and $\Lambda=\Lambda(r)$. When doing so we pretend to find
a solution which is still an exact solution of the equations
motion.
The corresponding improved equation of motion is \cite{Reuter:2004nx,Carroll:2004st}
(see also appendix \ref{AppendEOM})
\begin{eqnarray}
 G_{\mu\nu}=-g_{\mu\nu}\Lambda(r)+ 8 \pi G(r) T_{\mu \nu}
 -\Delta t_{\mu \nu}\quad,
\label{eom2}
\end{eqnarray}
with
\be
\Delta t_{\mu \nu}=G(r)\left(g_{\mu\nu}\Box-\nabla_\mu\nabla_\nu\right)\frac{1}{G(r)}\quad.
\ee
At first instance we are interested in spherically symmetric
solutions in regions where the classical
matter contribution $T_{\mu \nu}=0$.
As ansatz for the metric tensor we use
\be\label{ansatz}
ds^2=-f(r) dt^2+
1/f(r)dr^2+
r^2 d\theta^2
+r^2 \sin (\theta)d\phi^2
\ee
where
\be\label{fvonr}
f(r)=(1-2\frac{\Sigma G(r)}{r}-\frac{l(r)}{3}r^2).
\ee
With this ansatz the Einstein equations (\ref{eom2})
reduce to three independent differential equations for the
a priory unknown functions $G(r),\Lambda(r)$, and $l(r)$.
Please note that the constant $\Sigma$ would only be the mass of
the black hole, if $G$ and $\Lambda=l$ would be constants.
In the context of variable constants, however, $\Sigma$
is only an arbitrary constant with units of mass, which
can take arbitrary (even negative) values.

It was found that,
apart from the well known solutions, which
imply constant couplings, there exists a solution
with a non trivial $r$ dependence
\bea \label{Gvonr}
G(r)&=&-\frac{16 \pi c_2}{r-2 c_1}\quad,\\
\label{Lvonr}
\Lambda(r)&=&\frac{-1}{24 r (r-2 c_1)^2
c_1^4}\left\{-2 c_1 \left(c_1^2 \left(12 c_1^2+384 \Sigma \pi  c_2+c_3\right)+24 r^3 c_1^3 c_4+\right.\right.\\
\nonumber &&
\left.3 r^2 \left(384
\Sigma \pi  c_2+c_3-24 c_1^4 c_4\right)+6 r c_1 \left(-c_1^2-384 \Sigma \pi  c_2-c_3+8 c_1^4 c_4\right)\right)\\ &&
\left.+3 r
\left(r^2-3 r c_1+2 c_1^2\right) (384 \Sigma \pi  c_2+c_3) \ln[r]-3
r \left(r^2-3 r c_1+2 c_1^2\right) (384 \Sigma \pi  c_2+c_3) \ln[r-2 c_1]\right\}\quad,  \nonumber
\\ \label{llvonr}
l(r)&=&
c_4+\frac{1}{48 c_1^4}\left\{\frac{576 \Sigma \pi  c_1 c_2}{r-2 c_1}+\frac{8 c_1^3 \left(12 c_1^2+96 \Sigma \pi
c_2+c_3\right)}{r^3}+\frac{6 c_1^2 \left(12 c_1^2+192
\Sigma \pi  c_2+c_3\right)}{r^2}\right.\\ \nonumber&&
\left.
+\frac{6 c_1 (288 \Sigma \pi  c_2+c_3)}{r}-3 (384 \Sigma \pi  c_2+c_3)
\ln[r]+3 (384 \Sigma \pi  c_2+c_3) \ln[r-2
c_1]\right\}\quad.
\eea
Those solutions contain four constants of integration $c_i$. The role and the interpretation of those constants
in terms of different physical perspectives will be discussed throughout this paper.

One of the first questions one might ask for this new kind of solution
is whether it has the same singularity problem at the origin as the standard solution
or whether there exist parameter configurations for which the singularity can be avoided.
This can be checked by calculating the invariant tensor density which gives to lowest
order in $1/r$
\be
R_{\mu\nu\rho\sigma} R^{\mu\nu\rho\sigma} = \frac{144 c_1^4+9216 \Sigma \pi  c_1^2 c_2+147456 \Sigma^2 \pi ^2 c_2^2+24 c_1^2 c_3+768 \Sigma \pi  c_2 c_3+c_3^2}{27 c_1^2 r^6}+
{\mathcal{O}}(r^{-5}) \quad.
\ee
The divergence to this order can be avoided by choosing $c_2=\hat c_2$ with
\be
\hat c_2 =-\frac{12 c_1^2+c_3}{384 \Sigma \pi}\quad.
\ee
This choice removes a number of divergent terms such that the remaining next to next to next to leading singularity reads
\be
R_{\mu\nu\rho\sigma} R^{\mu\nu\rho\sigma}|_{\hat c_2} =\frac{2}{c_1^2 r^2}+
{\mathcal{O}}(r^{-1}) \quad.
\ee
If one further tries to remove this singularity, one is forced to take $\hat c_1= c_1 \rightarrow \infty$,
where the value for $c_2$ has to be chosen, before one takes the
limit in $c_1$.
This leads to a finite tensor density
\be
R_{\mu\nu\rho\sigma} R^{\mu\nu\rho\sigma}|_{\hat c_2, \hat c_1} = \frac{8}{3}c_4^2 \quad,
\ee
but this choice of the integration constants is actually almost trivial because corresponds to flat (anti) de Sitter space
\be
f(r)|_{\hat c_2, \hat c_1} =1-\frac{c_4}{3}r^2 \quad.
\ee

%%%%%%%%%%%%%%%%%%%%%%%%%%
\subsection{Solutions in the literature}

Since the equation of motion (\ref{eom2}) can be interpreted
as a special case of an $f(R)$ theory \cite{Nojiri:2006ri,Sotiriou:2008rp,Nojiri:2010wj}
(see \cite{Faraoni:2008mf} for a review)
one has to check,
whether the above solution has already been discussed
in the context of the more general $f(R)$ theories.
A class of solutions with constant curvature $R=R_0$
and perturbative expansions around the ``classical''
solutions has been discussed in \cite{delaCruzDombriz:2009et,Cognola:2005de}
and special solutions with conformal anomaly
have been discussed in \cite{Hendi:2012nj}.
Other perturbative solutions (assuming $g_{\mu \nu} \approx g^0_{\mu \nu} +h_{\mu \nu}$) to $f(R)$ theories
in their transition limit to general relativity can be found in \cite{Olmo:2006eh}.
Further variations of the classical solution have also been
studied in the context of Kerr black holes
\cite{Myung:2011we,Psaltis:2007cw,Barausse:2008xv}.
For an exact solution that assumes a finite
$1/(R-R\partial_R f(R))$ and constant curvature $R_0$,
 which generalizes to
charged, rotating black holes see
\cite{Larranaga:2011fv}.
None of the above solutions contains (\ref{Gvonr}-\ref{llvonr})
since here, $(R-R\partial_R f(R))=0$, $R$ is not constant,
and the solution is not a perturbation of a classical solution.
Thus, to our current knowledge (\ref{Gvonr}-\ref{llvonr})
has not been discussed in the context of $f(R)$ theories.

An other very similar approach is Brans Dicke
theory \cite{Brans:1961sx} where one allows for a varying gravitational
constant $\Phi(x)=1/G(x)$.
In difference to this approach, the Brans Dicke action
contains a kinetic term $\sim \omega \Phi_{,\mu}\Phi^{,\mu}$.
It the limit $\omega \rightarrow \infty$ this theory
corresponds to constant
coupling and standard general relativity.
Solutions of BD with the limit $\omega \rightarrow 0$
correspond to our ERG inspired approach.
Black hole solutions in pure Brans Dicke theory
have been discussed.
First black hole solutions for this theory have been found
by Brans \cite{Brans:1962zz}.
In \cite{Hawking:1972qk} it has been
shown that all BD black holes that have
constant $\Phi$ in the outside region and that
are static solutions correspond to
the known Schwarzschild, Kerr or Kerr-Newman solutions.
If one relaxes the condition of constant fields
further non-trivial solutions have been found
\cite{Kim:1998hc,Kim:2002mr}.
This solution does not apply here, since it has
$\Lambda=0$.
In \cite{Maeda:2011jj}
 BD black holes have been studied
in the Lifshitz context.
BD black hole solutions with a specific cosmological term $V(\Phi)$
have been found and discussed in
\cite{Gao:2004tu,Gao:2004tv,Gao:2005xv,Ghosh:2007jb}.
Based on those studies \cite{Sheykhi:2008tt,Sheykhi:2009zz}
discussed further generalizations of BD -AdS BHs.
By using dilaton black hole solutions with a cosmological
term $V(\bar \Phi)$ they construct via a conformal transformation
$\bar \Phi=(D-4)/(4\alpha) \ln \Phi$ the BD solution with
the cosmological term $V(\Phi)$.
This method, however, works only for larger
space-time dimensions $D>4$ since this conformal transformation
is ill defined for $D=4$. This reflects also in the
fact that the metric coefficients of those BD solutions diverge for four
space-time dimensions.

It is also instructive to do a comparison with an approximated black holes solution
of Brans Dicke theory given by Weinberg \cite{Weinberg:1972} (page 183 and page 247)
for the case that gives the Schwarzschild solution plus corrections in $1/r$
% parameters \alpha=\beta=1 and \gamma=1/2 with \omega=0
\be\label{epand1}
ds^2=(1-2\frac{MG}{r}+\frac{ M^2 G^2}{r^2}+\dots)dt^2
-(1+\frac{M G}{r}+\dots) dr^2
-r^2 d\theta^2 -r^2 \sin^2\theta d\phi^2\quad.
\ee
Comparing this approximation with $c_4=0$
to the series expansion in $1/r$ of the exact
solution one finds that one can at best fix the constants
$c_1$, and $c_{23}$ in combination with a time rescaling 
($t\rightarrow k_t t$), such that
\be\label{epand2}
ds^2=(1-2\frac{MG}{r}+\frac{M^2 G^2}{r^2}+\dots)dt^2
-(2+2\frac{M G}{r}+\dots) dr^2
-r^2 d\theta^2 -r^2 \sin^2\theta d\phi^2\quad.
\ee
Both expansions (\ref{epand1} and \ref{epand2})
do not agree in the constant coefficient of the $rr$
component. If one wants to demand a
regime where the exact solution has a constant factor
$1$ for both $g_{tt}$ and $g_{rr}$ one has
fix the constants for an expansion around $r\ll G_0\Sigma$,
as it is discussed in the following subsection.
This also can not be fixed by just varying the initial metric ansatz
by $g_{rr}\rightarrow k_r^2 g_{rr}$.
In terms of physical viability this is a severe problem, because
the above expansion for $c_4=0$ basically predicts that this new solution
would contradict all gravitational lensing effects with relativistic
trajectories. In order to address this concern we will first
ask, whether there exist special parameter choices (with $c_4\neq 0$) for which
the new solution approximates to the (for wide ranges of $r$)
well confirmed classical solution.

%%%%%%%%%%%%%%%%%%%%%%%%%%%%%%%%%%%%%%%%%%%%%%%%%%%%%%%%%%%%%%%%%%%%%%%%%%%%%%%

\subsection{A classical-like choice of parameters}
\label{subsecclaslike}
Given the problems for the $1/r$ expansion for $c_4=0$
we will now study the case $c_4\neq 0$ and
by trying to approximate the new solution to the standard solution
where
\be\label{fstandard}
f_s(r)=1-2 \frac{G_0 M_0}{ r}-r^2\frac{\Lambda_0 }{3}\quad.
\ee
As it can bee seen from equation (\ref{epand2}), not even reproducing
the standard result without cosmological constant is trivially possible.
Thus, one has to ask the question, whether there is any configuration
of the parameters ($c_1,c_2,c_3,c_4$) which is in agreement with
current experimental results, which basically confirm (\ref{fstandard}).
By constructing an approximation to the standard metric (\ref{fstandard})
 it will now be shown that such parameter choices exist.

\begin{figure}[hbt]
   \centering
%\centerline{\protect\vbox{\epsfig{file=Approx.eps,
%width=0.6\textwidth}}}
\includegraphics[width=10cm]{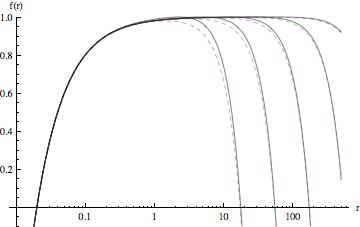}
  \caption{\label{figfr} Comparison of the classical metric coefficient (\ref{fstandard})
 with the solution (\ref{fvonrnew}) as dashed curves.
 The numerical values were chosen to be $M_0=0.01, G_0=1$ and from
 left to right
 $\Lambda_0=\{10^{-2}, 10^{-3}, 10^{-4},10^{-5},10^{-6}\}$.}
\end{figure}

An important property of the standard metric in the de Sitter case is that
it approximates to a plane flat space for large $r$, before it runs into
an other horizon at even larger distances.
Thus, we try to fix the constant $c_4$ such that there exists
a $r_m$ with
\be
f(r_m)=1 \quad{\mbox{and}} \quad f'|_{r_m}=0 \quad.
\ee
An analytic solution of those two combined conditions is possible, it gives
\bea
r_m&=& \sqrt{\frac{\tilde c_3+12 c_1^2}{3}}\quad,\\ \label{c4new}
c_{4,s}&=&\frac{12 c_1^2+\frac{4 \sqrt{3} \tilde{c}_3 c_1}{\sqrt{\tilde{c}_3+12 c_1^2}}+\frac{16 \sqrt{3} c_1^3}{\sqrt{\tilde{c}_3+12 c_1^2}}-\tilde{c}_3
\text{ln}[3]-\tilde{c}_3 \text{ln}\left[\tilde{c}_3+12 c_1^2\right]+2 \tilde{c}_3 \text{ln}\left[-6 c_1+\sqrt{3} \sqrt{\tilde{c}_3+12 c_1^2}\right]}{32
c_1^4}\quad,
\eea
where $\tilde c_3=c_3+382 \pi\Sigma c_2$.
The next step is to choose the constants $c_1$
and $c_3$ such that the two horizons of $f(r)$
approximate to the horizons of $f_s(r)$
\be
r_0\approx 2 G_0 M_0\quad{\mbox{and}} \quad
r_1 \approx \sqrt{\frac{3}{\Lambda_0}}\quad,
\ee
where we are interested in the cases $r_1\gg r_0$.
A numerical optimization to those conditions gives
\bea\label{c1new}
c_{1,s}&=&\frac{3^{2/3}}{4 (2 G_0 M_0 \Lambda_0^2)^{1/3}}\quad, \\ \label{c3new}
c_{3,s}&=&\frac{12 \cdot6^{2/3} G_0 (G_0 M_0)^{2/3} (-4 \Sigma+3 M_0) \Lambda_0^{4/3}-9 \cdot6^{1/3} \left(G_0 M_0 \Lambda_0^2\right)^{1/3}}{8 G_0 M_0 \Lambda_0^2}\quad.
\eea
By demanding that $G(r)\approx G_0$ for $r_1\gg r$ one can also fix the missing
constant
\be\label{c2new}
c_{2,s}=\frac{G_0}{32 \pi (2 G_0 M_0 \Lambda_0^2/9)^{1/3}}\quad.
\ee
Using the definitions (\ref{c4new}, \ref{c1new}, \ref{c3new}, \ref{c2new}),
the solution (\ref{fvonr}, \ref{Gvonr}, \ref{Lvonr}) can be written as
\begin{small}
\bea
f(r)&=&\frac{1}{9\cdot 6^{2/3} r \left(\frac{G_0 M_0}{\Lambda_0}\right)^{1/3}}
\left(9\cdot 6^{2/3} r \left(\frac{G_0 M_0}{\Lambda_0}\right)^{1/3}\right.-144\cdot 6^{1/3} G_0^2 M_0^2 r^2 \Lambda_0-18\cdot 6^{2/3} \left(\frac{G_0 M_0}{\Lambda_0}\right)^{4/3}
\Lambda_0\nonumber \\ && \label{fvonrnew}
+48 \sqrt{6} (G_0 M_0)^{5/3} r^3 \Lambda_0^{4/3}+36 r^2 \left(G_0^2 M_0^2 \Lambda_0\right)^{1/3}-108
r \left(G_0^5 M_0^5 \Lambda_0\right)^{1/3}-8\cdot 6^{1/6} r^3 \left(G_0 M_0 \Lambda_0^2\right)^{1/3}\\ \nonumber&&
+12 \cdot6^{1/3}
G_0 M_0 r^3 \Lambda_0 (\text{ln}[3]+1)-48 \cdot6^{2/3} (G_0 M_0)^{7/3} r^3 \Lambda_0^{5/3} \text{ln}[3]\\ \nonumber&&
+12\cdot 6^{1/3}
G_0 M_0 r^3 \left(1-4 \cdot6^{1/3} (G_0 M_0)^{4/3} \Lambda_0^{2/3}\right) \Lambda_0 \text{ln}\left[\frac{9\cdot
3^{2/3} \left(\frac{G_0 M_0}{\Lambda_0}\right)^{2/3}}{2^{1/3}r^2}\right]\\ \nonumber&&
\left.
+\left(96 \cdot6^{2/3} (G_0 M_0)^{7/3} r^3 \Lambda_0^{5/3}
-24\cdot 6^{1/3} G_0 M_0 r^3 \Lambda_0\right)
\text{ln}\left[\frac{3 \left(-6^{2/3}+2\cdot 6^{5/6} \left(G_0^2 M_0^2 \Lambda_0\right)^{1/3}\right)}{
4 r \left(G_0 M_0 \Lambda_0^2\right)^{1/3}-6^{2/3}
}\right]\right)\quad,
\eea
\end{small}
\be\label{Gvonrnew}
G(r)=\frac{6^{2/3} G_0}{6^{2/3}-4 r (G_0 M_0 \Lambda_0^2)^{1/3}}\quad,
\ee
\begin{small}
\bea\label{Lvonrnew}
\Lambda(r)&=&
\frac{-4 2^{1/3} \Lambda_0}{3\cdot 3^{2/3} (G_0 M_0)^{2/3} r \left(6^{2/3}-4 r \left(G_0 M_0
\Lambda_0^2\right)^{1/3}\right)^2}
\left\{\frac{6}{6^{\frac{1}{6}}} 36G_0^2 M_0^2 r \Lambda_0^{2/3}\right.-6 \sqrt{6} (G_0 M_0)^{2/3} r
+12\cdot 6^{5/6} G_0 M_0 r^2 \Lambda_0^{2/3}\\
\nonumber &&
+864 (G_0 M_0)^{8/3} r \Lambda_0+36 (G_0 M_0)^{5/3} r^2
\Lambda_0-576\cdot 6^{1/3} G_0^3 M_0^3 r^2 \Lambda_0^{5/3}+192
\sqrt{6} (G_0 M_0)^{8/3} r^3 \Lambda_0^2\\
\nonumber &&
-432\cdot 6^{1/6} G_0 M_0 r^2 (G_0 M_0 \Lambda_0)^{4/3}-32
\cdot 6^{1/6} r^3 (G_0 M_0 \Lambda_0)^{4/3}-24\cdot 6^{2/3} r \left(G_0^4 M_0^4 \Lambda_0\right)^{1/3}-18\cdot 6^{2/3}
\left(G_0^7 M_0^7 \Lambda_0\right)^{1/3}\\
\nonumber &&
-216 (G_0 M_0)^{8/3} r \Lambda_0 \text{ln}[3]-108 (G_0
M_0)^{5/3} r^2 \Lambda_0 \text{ln}[3]+432\cdot 6^{1/3} G_0^3 M_0^3 r^2 \Lambda_0^{5/3} \text{ln}[3]+48\cdot 6^{1/3}
G_0^2 M_0^2 r^3 \Lambda_0^{5/3} (\text{ln}[3]+1)\\
\nonumber &&
-192\cdot 6^{2/3} (G_0 M_0)^{10/3} r^3 \Lambda_0^{7/3} \text{ln}[3]+9
\cdot 6^{2/3} G_0 M_0 r (G_0 M_0 \Lambda_0)^{1/3} \text{ln}[3]\\
\nonumber &&
-6 r \left(-36 (G_0 M_0)^{5/3} r \Lambda_0
+8 G_0^2 M_0^2 \left(-9 (G_0 M_0)^{2/3}+2\cdot 6^{1/3} r^2 \Lambda_0^{2/3}\right) \Lambda_0+3\cdot 6^{2/3}
\left(G_0^4 M_0^4 \Lambda_0\right)^{1/3}\right.\\
\nonumber &&
\left.-16\cdot 6^{1/3} G_0^3 M_0^3 r \Lambda_0^{5/3} \left(-9+4\cdot 6^{1/3}
r \left(G_0 M_0 \Lambda_0^2\right)^{1/3}\right)\right) \text{ln}[r]\\
\nonumber &&
-3 G_0 M_0 r \left(72 (G_0 M_0)^{5/3}
\Lambda_0+36 (G_0 M_0)^{2/3} r \Lambda_0-144\cdot 6^{1/3} G_0^2 M_0^2 r \Lambda_0^{5/3}-16\cdot 6^{1/3}
G_0 M_0 r^2 \Lambda_0^{5/3}\right.\\
\nonumber &&
\left.-3\cdot 6^{2/3} (G_0 M_0 \Lambda_0)^{1/3}+64\cdot 6^{2/3} r^2 (G_0 M_0 \Lambda_0
)^{7/3}\right) \text{ln}\left[\frac{9\cdot 3^{2/3} \left(\frac{G_0 M_0}{\Lambda_0}\right)^{2/3}}{2^{1/3}}\right]\\
\nonumber &&
+\left(
432 (G_0
M_0)^{8/3} r \Lambda_0 
-18 \cdot6^{2/3} G_0 M_0 r (G_0 M_0 \Lambda_0)^{1/3} 
216 (G_0 M_0)^{5/3} r^2 \Lambda_0 
+384\cdot 6^{2/3} (G_0 M_0)^{10/3} r^3 \Lambda_0^{7/3}
\right.\\
\nonumber &&\left.
-\left.864\cdot6^{1/3} G_0^3 M_0^3 r^2 \Lambda_0^{5/3} 
-96\cdot 6^{1/3} G_0^2 M_0^2 r^3
\Lambda_0^{5/3}\right)
\text{ln}\left[\frac{3 \left(-6^{2/3}+2\cdot 6^{5/6} \left(G_0^2 M_0^2 \Lambda_0\right)^{1/3}\right)}{
4 r \left(G_0 M_0 \Lambda_0^2\right)^{1/3}-6^{2/3}
}\right]\right\}\quad.
\eea
\end{small}
One can see that now there appears only the physical mass $M_0$
and not the parameter $\Sigma$  from the initial ansatz in this set of solutions.
The physical meaning of $M_0$ can be verified by expanding
(\ref{fvonrnew}) for small $\Lambda_0$
\be
f(r)=1-2\frac{G_0 M_0}{r}+{\mathcal{O}}\left(\Lambda_0^{2/3}\right)\quad.
\ee
This special choice of the parameters ($c_{1,s},c_{2,s},c_{3,s},c_{4,s}$) is not exactly the most compact
but it shows that the classical result $f_s(r)$ with the parameters
($\Lambda_0, M_0, G_0$) can be approximated very well by the
exact solution $f(r)$ with variable $G(r)$ and $\Lambda(r)$.
As it can be seen, from figure \ref{figfr}, $f(r)$ and $f_s(r)$ are practically indistinguishable
for very small values of $\Lambda_0$.
%
%%%%%%%%%%%%%%%%%%%%%%%%%%%%%%%%%%%%%%%%%%%%%%%%%%%%%%%%%%%%%%%%%%%%%%%%%%%%%%%%%%
%%%%%%%%%%%%%%%%%%%%%%%%%%%%%%%%%%%%%%%%%%%%%%%%%%%%%%%%%%%%%%%%%%%%%%%%%%%%%%%%%%
Since the classical form of (\ref{fstandard}) has been reproduced
by various experiments at different scales, and since the special
choice of $c_4$ allows to approximate very well to this result,
it can be concluded that the solution (\ref{fvonr})
allows also for parameter choices which are also in agreement with current
experimental limits and observations. The question
to which extend observational data such as dark matter effects
can be incorporated in the current solution 
(in the spirit of \cite{Reuter:2004nx,Reuter:2004nv,Rodrigues:2009vf,Tye:2010an,Bonanno:2010bt}) will be subject
of future studies \cite{DaviOliverBen}.
The above choice of parameters is most likely not the only one
that achieves this goal, but the point is that it shows by construction
that physically viable parameter choices are perfectly possible.

%\pagebreak
%%%%%%%%%%%%%%%%%%%%%%%%%%
\section{Coupling Flow}\label{SecRun}

\subsection{The induced coupling flow}
\begin{figure}[hbt]
   \centering
%%\centerline{\protect\vbox{\epsfig{file=fvsr_d2_tt001_Mf1.eps,
%%width=0.6\textwidth}}}
\includegraphics[width=10cm]{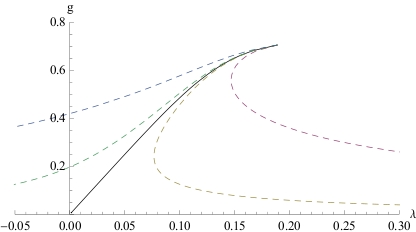}
  \caption{\label{figflowUV} Schematic flow of the scale dependent
  couplings $\lambda_{U}(r)$ and $g_{U}(r)$ for $g^*_U=0.707$,
  $\lambda^*_U=0.193$, $g_I=2.5$, and
  $G_0=\Sigma=1$.
    The different curves correspond to
  $l_I=\{ -0.05,\,-0.005,\,0,\,0.005,\,0.05\}$ which are depicted from left
  to right in $\{$blue,~green,~black,~orange,~red$\}$.
  }
\end{figure}

When discussing the flow of scale dependent couplings
like ($G, \Lambda$)
one usually does so for the dimensionless couplings ($g, \lambda$).
Therefore, the free parameters that describe
this flow, should also be adimensional.
Hence one defines a set of four
dimensionless parameters parameters
$\{\lambda^*_U,\,l_I,\, g^*_U,\, g_I\}$
instead of the dimensionfull parameters
$\{c_1,\,c_2,\,c_3,\,c_4\}$
\be\label{replFP}
\left.
\begin{array}{ccc}
 c_1&=&-\frac{g_I}{2 g^*_U \Sigma} \\
 c_2&=&-\frac{g_I}{16 \Sigma^3 \pi}\\
 c_3&=&
 \frac{3 g_I (8 g^{*3}_U - g_I g^*_U + 2 g^{2}_I \lambda^*_U)}
 {g^{*3}_U \Sigma^2}\\
 c_4&=&-\frac{\Sigma^2l_I}{2}
\end{array}
\right\}
\quad
{\text{or inversely}}
\quad
\left\{
\begin{array}{ccc}
 \lambda^*_U&=& -\frac{12 c_1^2 + c_3+ 384 c_2 \Sigma \pi}{48 c_1^3 \Sigma }\\
l_I&=&-\frac{2 c_4}{\Sigma^2}\\
 g^*_U &=& \frac{8 c_2 \Sigma^2 \pi}{c_1}\\
 g_I &=&-16 c_2 \Sigma^3 \pi
\end{array}
\right\}\quad.
\ee
With this reparametrization
the metric solution reads
\bea\label{fvonrnew2}
f(r)&=&
\frac{1}{6
g_I^2 g^{*2}_U  \Sigma  r}
\left\{g_I \left(-6 g^{*3}_U  \Sigma ^2 r^2+4 g_I^3 \lambda^{*}_U -6 g_I^2 g^{*}_U  \Sigma  r \lambda^{*}_U +g_I
g^{*2}_U  \Sigma  r \left(6+\Sigma ^2 r^2 l_I+12 \Sigma  r \lambda^{*}_U \right)\right)\right.\\ \nonumber
&&\left.
+6 g^{*3}_U  \Sigma ^3 r^3 (g^{*}_U -2 g_I \lambda^{*}_U ) 
\text{Log}\left[\frac{g_I}{g^{*}_U  \Sigma  r}+1\right]\right\}\quad.
\eea
One observes that the remaining
dimensionfull parameters are $r$ and $\Sigma$, which
only appear in dimensionless pairs $r \Sigma$ in (\ref{fvonrnew}).
Now that convenient parameters have been defined
one can define the dimensionless couplings.
In order to make he couplings dimensionless one has
to multiply them with dimension-full quantities that describe
the physical system. In our case the two dimension-full
quantities that can vary and that describe the physical system are $\Sigma$ and $r$.
In principle, every adequate power of those two quantities can do the job,
thus one can write
\bea\label{adimg}
g(r)&=& G(r)\frac{\Sigma^2}{(\Sigma r)^a}\\ \label{adiml}
\lambda(r)&=&-\Lambda(r)\frac{(\Sigma r)^c}{\Sigma^2}\quad,
\eea
where $a$ and $c$ are numbers that determine the
respective impact of $r$ and $\Sigma$ ``adimensionalization!"
and the minus sign in (\ref{adiml}) is pure convention.
The constants $a$ and $c$ are crucial for the expected fixed point behavior
of the adimensional coupling flow.
Only for certain values there exists
a well behaved, non-trivial fixed point.
Motivated from the results in the ERG approach, which will be
introduced in the following section, 
one demands the existence of a non-trivial UV fixed point.
This UV fixed point exists for both
couplings for the values
$a=0$ and $c=+1$
\bea\label{UVFP}
g_{U}(r)&=& G(r) \Sigma^2 \\ \nonumber
\lambda_{U}(r)&=& -\Lambda(r) \frac{r}{\Sigma} \quad.
\eea
The values of the UV fixed points are
\bea\label{UVFPlim}
g_{U}(r\rightarrow 0)&=&g_{U}^*\\ \nonumber
\lambda_{U}(r\rightarrow 0)&=&\lambda_{U}^* \quad.
\eea
The limits in (\ref{UVFPlim})
show that two of the new dimensionless parameters, 
are such that they represent the numerical
value of the corresponding UV fixed points. Therefore, part of the possible
physical results are already encoded in the numerical value of this fixed point.
The choice (\ref{UVFP}) is further interesting in the sense that
in the adimensional couplings only appear the adimensional quantities
$\lambda_U^*,\; l_I,\; g^*_U,\; g_I$, and $ (r \cdot \Sigma)$.
With those parameters the adimensional gravitational couplings read
\begin{equation}\label{adimgU}
g_U(r)= \frac{{g^{*}_{U}}}{ 1 +\frac{ g^{*}_{U}}{ g_{I} } \Sigma r}\quad.
\end{equation}
and
\bea
\lambda_U(r)&=&
\frac{1}{2 g_I^2 (g_I+g^{*}_U \Sigma  r)^2}\left\{
g_I \left(g_I^3 (\Sigma  r l_I+2 
\lambda^{*}_U)-12g^{*3}_U \Sigma ^2 r^2+3 g_I^2g^{*}_U \Sigma  r (\Sigma  r
l_I+8 \lambda^{*}_U)\right)\right.\\ \nonumber
&+& \left.
g_I^2g^{*2}_U \Sigma  r \left(2 \Sigma ^2 r^2 l_I-11+24 \Sigma  r \lambda^{*}_U\right)
+6 g^{*}_U \Sigma  r \left(g_I^2+3 g_Ig^{*}_U \Sigma  r+2g^{*2}_U \Sigma ^2 r^2\right) 
(g^{*}_U-2 g_I \lambda^{*}_U) \ln
\left[\frac{g_I}{ g^{*}_U \Sigma r}+1\right]\right\}.
\eea
It is straight forward to depict the flow of this
UV fixed point scenario $a=0$ and $c=1$.
The corresponding flow is shown in figure \ref{figflowUV},
where the numerical values of $g_U^*$ and $\lambda_U^*$ where taken from \cite{Groh:2010ta}.
%%%%%%%%%%%%%%%%%%%%%%%%%%
\subsection{Comparing to the ERG flow}

\begin{figure}[hbt]
   \centering
%%\centerline{\protect\vbox{\epsfig{file=fvsr_d2_tt001_Mf1.eps,
%%width=0.6\textwidth}}}
\includegraphics[width=10cm]{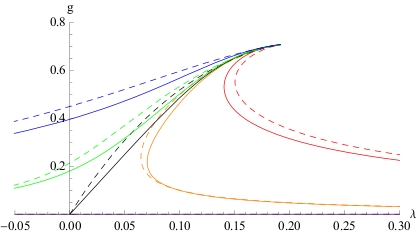}
  \caption{\label{figflowUV2} Flow of the scale dependent
  couplings due to the ERG result from equation 
  (\ref{lambdaERGg}) which are depicted as dashed lines, and
 by using the parameters
$g^*_U=0.707$,  
  $\lambda^*_U=0.193$, and $G_0=1$ with
  $\Lambda_0=\{ -0.1,\,-0.01,\,0,\,0.01,\,0.1\}$ ($\{$blue,~green,~black, orange,~red$\}$). 
 This flow is compared to the black hole induced flow
  $\lambda_{U}(r)$ and $g_{U}(r)$ as solid lines for 
  the same parameters with the identifications 
  $\Lambda_0=l_I \cdot \Sigma^2$ and $G_0=g_I/\Sigma^2$
  with the additional choice of $g_I=2.5$
  and $\Sigma=1$.
  }
\end{figure}

In the context of ERG induced coupling flows
the dimensionless couplings are defined by the use
of the energy scale $k$
\be\label{Gvong}
 g(k)=k^{2} G(k)\quad, \quad
\lambda(k)=\frac{\Lambda(k)}{k^{2}}\quad.
\ee
Those running couplings
and the corresponding fixed points have been repeatedly calculated numerically \cite{Dou:1997fg,Souma:1999at,Reuter:2001ag,Litim:2003vp,Fischer:2006fz,Niedermaier:2006wt,Litim:2008tt,Narain:2009gb,Niedermaier:2009zz,Groh:2010ta,Rechenberger:2012pm}.
In order to obtain a tractable analytic solution for the running couplings (\ref{Gvong})
we will use a similar approximation procedure as it was used in \cite{Koch:2010nn}.
According to \cite{Litim:2003vp}
the flow equations   are given by:
\begin{eqnarray}
\partial_{t}g(k) &=& \beta_{g}(\lambda_{k},g_{k})=[d-2+\eta(k)]g(k)
\nonumber \\
\partial_{t}\lambda(k)&=& \beta_{\lambda}
(\lambda_{k},g_{k}),
\end{eqnarray}
where $t= \ln k/\Lambda$ and $\eta$ is the anomalous dimension and  the beta functions  are for $g$
\be
\partial_t g=\beta_g=\frac{(-2+d) P_2 g(k)}{P_2+4 (2+d) g(k)}\quad,
\ee
and for $\lambda$
\be
\partial_t \lambda= \beta_\lambda=\frac{P_1}{P_2+4 (2+d) g(k)}\quad.
\ee
with
\bea
P_1&=&d (2+d) g(k) (-3+d-16 g(k)+8 d g(k))+4 \left(-1+10 d g(k)+d^2 g(k)-d^3 g(k)\right) \\ \nonumber
&&\lambda (k)+4 \left(4-10 d g(k)-3 d^2 g(k)+d^3 g(k)\right) \lambda
(k)^2-16 \lambda (k)^3\quad,
\eea
and
\be
P_2=2+8 \left(-d g(k)-\lambda (k)+\lambda (k)^2\right)\quad.
\ee
Expanding those beta functions for small values of the couplings 
$(g,\lambda \ll 1)$
and for four 
space-time dimensions one gets
\be\label{betagapprox}
\beta_g=g(k)(2 -24 g(k))
\ee
and
\be\label{betalapprox}
\beta_\lambda=12 g(k)-2 \lambda (k)\quad.
\ee
The approximated beta function (\ref{betagapprox}) can be integrated to
\be \label{gERG}
g_{ERG}(k)=\frac{ g_U^{*}}{1+\frac{G_0 k^2}{g_U^{*}}}\quad.
\ee
Using (\ref{gERG}) one can also integrate (\ref{betalapprox})
giving
\bea\label{lambdalit}
\lambda(k)_{ERG}&=&
\lambda^{*}_U+\frac{1}{k^2} \Lambda_0
-\frac{g^*_U \lambda^*_U}{G_0 k^2}
\text{Log}\left[ \left(1+G_0\frac{k^2}{g_U^{*}}\right)\right]\quad.
\eea
Note that the fixed points for this  flow equations are the Gaussian fixed point ($(\lambda^*,g^*)=(0,0)$)
  and in this approximation the UV non Gaussian fixed point  with $\lambda^*_U=1/2$
and $g^*_U=1/12$.
The values of the non Gaussian fixed point were replaced in the solution by their symbols
$g^*_U$ and $\lambda^*_U$, which will subsequently be treated as free parameters. The relation between $g$ and $k^2$ (\ref{gERG}) can be inverted in order to express equation (\ref{lambdalit})
in terms of $g_{ERG}$, giving
\be\label{lambdaERGg}
\lambda_{ERG}(g)=
\lambda^*_{U} +
\frac{ 1}{ g }\left(  \Lambda_{0} G_{0} (1-g /g^*_{U} ) -\lambda^*_{U}  g^*_{U} (1- g /g^*_{U} ) \ln\left[\frac{1}{1-g /g^*_{U} }\right]\right)
\ee
This result can be plotted and compared to the coupling flow from figure
\ref{figflowUV}, the resulting graphical comparison of the two flows
is shown in figure \ref{figflowUV2}.

Since the graphical similarity of the black hole induced flow
and of the analytic ERG flow in figure \ref{figflowUV2} is quite striking
we will now proceed with an analytical comparison.

For the adimensional gravitational constant one finds that
the black hole induced $g_U$ (\ref{adimgU}) and the ERG result $g_{ERG}$
(\ref{gERG}) are exactly identical if one uses the scale setting
\be\label{setscaler}
r \equiv\frac{g_I}{k^2 G_0 \Sigma}\quad.
\ee
This scale setting result is interesting since intuitively one might have
expected something like $k\sim 1/r$.
Please note that this scale setting definition still leaves $g_I$ arbitrary.
The next step is to compare the couplings $\lambda_{ERG}$ (\ref{lambdalit})
and $\lambda_U$ (\ref{adiml}).
By using the scale setting (\ref{setscaler}) 
one finds
\begin{small}
\bea\label{lambdaBH}
\lambda_U(k)&=&
\frac{1}
{2 g_I  k^2 \left(g_U^{*}+ G_0 k^2\right)^2}\\ \nonumber
&&
\left\{G_0^2 g_Ik^4 \left(\frac{g_I l_I}{G_0}+2 k^2 \lambda_U^{*}\right)-12 g_U^{*3} k^2+3G_0 g_I
g_U^{*} k^2 \left(\frac{g_I l_I}{G_0}+8 k^2 \lambda_U^{*}\right)+g_U^{*2} \left(-11G_0 k^4+2 g_I\left(\frac{g_I l_I}{G_0}+12
k^2 \lambda_U^{*}\right)\right)\right.\\ \nonumber
&&
\left.+6 \frac{g_U^{*}}{G_0} \left(2 g_U^{*2}+3G_0 g_U^{*} k^2+ G_0^2
k^4\right) (g_U^{*}-2 g_I\lambda_U^{*}) \text{Log}\left[ \left(1+G_0\frac{k^2}{g_U^{*}}\right)\right]\right\}
\eea
\end{small}
The limit $\lim_{k\rightarrow \infty}=g_I l_I/G_0$ suggests the identification 
$l_I\equiv \frac{\Lambda_0 G_0}{g_I}$.
Apparently, (\ref{lambdaBH}) is not identical to (\ref{lambdalit}), but the question 
is whether and to which extent both are similar.
Since equation (\ref{lambdalit}) is an analytic approximation which is assumed to be
best close to a small valued fixed point it is instructive to compare (\ref{lambdalit})
and (\ref{lambdaBH}) in the UV regime for large values of $k^2$.
For the comparison we separate (\ref{lambdaBH}) in a logarithmic and a non-logarithmic
part and perform a Taylor expansion of the coefficients to lowest
order in $(1/k^2)$, $\lambda^*_U$, and $g^*_U$, which is analogous
to the expansion that was used when deriving $\lambda_{ERG}(k)$. This gives
\bea\label{lambdaBHexp}
\lambda_U(k)|_{UV}&=&
\lambda^{*}_U+\frac{1}{k^2} \frac{g_I l_I}{2G_0}\\ \nonumber
&&
- \frac{g^{*}_U \lambda^*_U}{G_0k^2}
\frac{(6 g_I -3 \frac{g^{*}_U}{\lambda^*_U} )  }{ g_I}
\text{Log}\left[ \left(1+G_0\frac{k^2}{g_U^{*}}\right)\right] \quad.
\eea
One observes that the approximated 
black hole induced function (\ref{lambdaBHexp}) has the same 
functional structure as the approximated ERG function (\ref{lambdalit}).
Even more, by choosing the remaining free constants to be
$l_I=2 G_0 \Lambda_0/g_i$ and $g_I=3 g^*_U/(5 \lambda^*_U)$,
the matching is exact.
The approximated BH induced cosmological constant (\ref{lambdaBHexp})
is then identical to the
(approximated) ERG function  (\ref{lambdalit}).
Please note that due to the UV approximation the infrared limit $k\rightarrow 0$ of  
(\ref{lambdaBHexp}) is a factor of two different from the infrared limit of 
the complete expression (\ref{lambdaBH}).

%%%%%%%%%%%%%%%%%%%%%%%%%%%%%%%%%%%%%%%%%%%%%%%%%%%%%

%%%%%%%%%%%%%%%%%%%%%%%%%%%%%%%%%%%%%%%%%%%%%%%%%%%%%%%%%%%%%%%%%%%%%%%%%%%%%%%%%%
%%%%%%%%%%%%%%%%%%%%%%%%%%%%%%%%%%%%%%%%%%%%%%%
\subsection{Anomalous  dimension and product of fixed points}

\begin{figure}[hbt]
   \centering
%%\centerline{\protect\vbox{\epsfig{file=fvsr_d2_tt001_Mf1.eps,
%%width=0.6\textwidth}}}
\includegraphics[width=10cm]{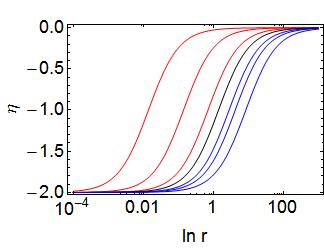}
\caption{{\protect\footnotesize 
\label{anom123}
Anomalous dimension  as a function of $r$, for   $\Sigma = 1$, $g^*_U=0.707$,  with
$g_I= (0.01, \,0.1, \,0.5)$ in red, $g_I=1$ in black, and 
$g_I=(1.5,\, 2.5,\, 3.5)$ in blue.
}}
\end{figure}

One  relevant point in the discussion of the running parameter is  the behavior of the anomalous dimensions in the UV or IR region.
The anomalous dimension $\eta$ is connected to the previously defined
beta function $\beta_g$
\begin{eqnarray}
\partial_{t}g(k) &=& \beta_{g}(\lambda_{k},g_{k})=[d-2+\eta(k)]g(k)
\end{eqnarray}
where $t= \ln k/\Lambda$.
By using (\ref{Gvong}) $\eta$ can be written as
\begin{equation}
\eta(k)=  -2 +\frac{1}{g(k)}\partial_{t}g(k)
\end{equation}
The conditions for the  existence of  non trivial fixed points in the UV   limit,
non Gaussian fixed points,  is   that: $\beta_{g}=0$ and $\beta_{\lambda}=0$. With those conditions  the UV behavior of  the anomalous dimension  is given by   $k \mapsto \infty$   ; $\eta_{k}\mapsto  -(d-2)$.
In order to study  the behavior of the anomalous dimension in the IR  region  we select one of  the  trajectories which connects the UV non Gaussian fixed point and run to the  IR region as it is done in figure \ref{anom123}. 
One can see from the figure that 
the trajectories have a classical behavior in the IR 
where the  anomalous dimension goes to zero, while the 
anomalous dimension smoothly goes to the fixed point value in the UV:
\begin{itemize}
 \item  In the  IR  limit $k \mapsto 0$; $\eta_{k}\mapsto  0 $
  \item  In the  UV  limit $k \mapsto \infty$   ; $\eta_{k}\mapsto-(d-2)$
  \end{itemize}

Since in the previous discussion it has been shown that $g_{ERG}(k)$
and $g_U(r)$ are exactly equivalent due to the scale setting condition (\ref{setscaler})
it is sufficient to work with one of the two, for example  $g_{U}(r)$ (\ref{adimgU}). Considering this result we find that  anomalous  dimension has the form:
\begin{equation}
\eta(r)= -2 -2   \frac{\partial \ln  G(r) } {\partial \ln r} = -2 +2   \frac{r/g_I } {\frac{ {1}}{ g^{*}_{U}\Sigma}+r/g_I}.
\end{equation}
One can see nicely that $r$ and $g_I$ only appear in pairs $r/g_I\sim /k^2$
which explains that the different values of $g_I$ in figure \ref{anom123} 
actually correspond to a rescaling of $r$ in the same function $\eta(r)$.

%%%%%%%%%%%%%%%%%%%%%%%%%%%%%%%%%%%%%%%%%%%%%%%

\begin{figure}[hbt]
   \centering
%%\centerline{\protect\vbox{\epsfig{file=fvsr_d2_tt001_Mf1.eps,
%%width=0.6\textwidth}}}
\includegraphics[width=10cm]{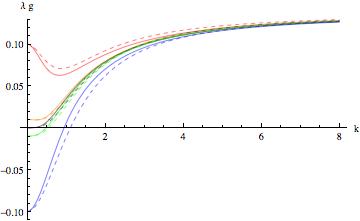}
  \caption{\label{figltimesg} Product of the scale dependent
  couplings due to the ERG result from equation 
  (\ref{lambdaERGg}) which are depicted as dashed lines, and
 by using the parameters
$g^*_U=0.707$,  
  $\lambda^*_U=0.193$, and $G_0=1$ with
  $\Lambda_0=\{ -0.1,\,-0.01,\,0,\,0.01,\,0.1\}$ ($\{$blue, green, black, orange, red$\}$). 
 This is compared to the black hole induced product
  $\lambda_{U}(k)g_{U}(k)$ as solid lines for 
  the same parameters with the identifications 
  $\Lambda_0=l_I \cdot \Sigma^2$ and $G_0=g_I/\Sigma^2$
  with the additional choice of $g_I=2.5$
  and $\Sigma=1$.
  }
\end{figure}

In the variety of ERG calculations it turned out that even
though the values of the fixed points $g_{ERG}^*$ and $\lambda_{ERG}^*$
are scheme dependent, the product $g_{ERG}^*\cdot \lambda_{ERG}^*$
is rather robust throughout the different calculations.
Therefore, figure (\ref{figltimesg}) shows the product
$g_{ERG}\cdot \lambda_{ERG}$ compared to the product
$g_{U}\cdot \lambda_{U}$ as a function of $k$.

One can see from figure \ref{figltimesg} that also the ERG result
and the black hole induced results for $\lambda \cdot g$
are in good agreement.

%\begin{figure}[htb]
%\begin{minipage}[b]{.39\linewidth}
%\begin{center}
%\centering\epsfig{file=anom123.eps,width=\linewidth}
%\end{center}
%\end{minipage}
%\vspace{0.1cm}
%\end{figure}
%%%%%%%%%%%%%%%%%%%%%%%%%%%%%%%%%%%%%%%%%%%%%%%%%%%%%

%%%%%%%%%%%%%%%%%%%%%%%

%%%%%%%%%%%%%%%%%%%%%%%%%%%%%%%%%%%%%%%%%%%%%%%%%%%%%55
%
%%%%%%%%%%%%%%%%%%%%%%%%%%%%%%%%%%%%%%%%%5
%\pagebreak
\section{Summary and Conclusion}
\label{SecConcl}

In this paper we have studied the possibility of a scale dependent
gravitational coupling $G(r)$ and cosmological coupling $\Lambda(r)$.
Such scale dependent couplings have to be studied in the context of improved
equations of motion~(\ref{eom2}). 
We asked the question of how the scale dependent couplings would have to look like
in order to permit the most simple spherical symmetric metric solution with $g_{tt}=-1/g_{rr}$.
Solving the equations of motion lead to a non-trivial metric $g_{tt}(r)$ with
non-trivial functions for $G(r)$, and $\Lambda(r)$.
This solution contains four constants of integration. 
Since a naive expansion with with one of the constants zero lead to unphysical
predictions (\ref{epand2}) for large radii, we showed the existence
of parameter choices where this problem does not exist (see section \ref{subsecclaslike}).

From the functional form of the dimension-full coupling constants $(G(r), \, \Lambda(r))$,
we defined the dimensionless coupling constants $(g_U(r), \, \lambda_U(r))$
with a fixed point in the regime of small radii.
The flow of those dimensionless couplings was then compared to the 
(approximated) functional
form of the running couplings of the ERG approach $(g_{ERG}(k), \, \lambda_{ERG}(k))$.
After a proper scale setting $k=k(r)$ (\ref{setscaler}) it was found that there 
exists an exact equivalence between $g_U(r) \;\& \; g_{ERG}(k)$
and a structural correspondence between  $\lambda_U(r)\;\&\; \lambda_{ERG}(k)$.
By approximating the black hole induced result 
$\lambda_U(r)\rightarrow \lambda_U(r)|_{UV}$
to the same order as the ERG result $\lambda_{ERG}(k)$ one even finds an
exact agreement between $ \lambda_U(r)|_{UV}$ and $\lambda_{ERG}(k)$.
Finally, the behavior of the anomalous dimension
and of the product of the two couplings was discussed.

Given the good qualitative and quantitative agreement between
the actual ERG result and the black hole induced results,
one is tempted to believe that the solution (\ref{fvonr}, \ref{Gvonr}, and \ref{Lvonr})
is actually a selfconsistent and good approximation to a still unknown
complete solution of the ERG improved equations of motion.
The solution is defined for all scales, but its correspondence to
the ERG black hole is expected
to be best for small values of $g^*_U$ and $\lambda^*_U$ and for large energy
scales $k$.
This similarity (and approximate correspondence)
was taken as an unexpected surprise and is the main result of this study.

\section*{Acknowledgements}
The work of B.K.\ was supported proj.\ Fondecyt 1120360
and anillo Atlas Andino 10201;
The work of C.C.\ was supported proj.\ Fondecyt 1120360 and DGIP grant  11.11.05. 
%%%%%%%%%%%%%%%%%%%%%%%%%%%%%%%%%%%%%%%%%%%55
%\bibliographystyle{h-physrev.bst}
%\bibliographystyle{apsrev}
%\bibliography{../bibfile}
%\begin{thebibliography}{10}

\pagebreak
%%%%%%%%%%%%%%%%%%%%%%%%%%%%%%%%%%%%%%%%%%%%%%%%%%%%%55
\begin{appendix}
\section{Complementary Material}\label{SecAppend}

%%%%%%%%%%%%%%%%%%%%%%%%%%%%%
\subsection{Improved action and improved equation of motion}
\label{AppendEOM}

Coupling the Einstein-Hilbert action to matter with scale dependent
couplings $\Lambda(k)$ and $G_g$ gives the improved action
\begin{eqnarray}
 S[g]&=&\int_{M}  d^4 x
 \sqrt{-g}\left(\frac{R-2\Lambda_{ k}}{16\pi G_{
k}}+\mathcal{L}_m\right)
-\frac{1}{8 \pi}\int_{\partial M} d^3x \sqrt{-h}\frac{K}{G(k)}
\quad. \label{action}
\end{eqnarray}
The Gibbons-Hawking boundary term with the
trace of the extrinsic curvature $K$, can become
relevant for the consistency of
solutions that are not asymptotically flat such as (A)dS.
The equations of motion for the metric field in (\ref{action}) are
\begin{eqnarray}
 G_{\mu\nu}=-g_{\mu\nu}\Lambda_{k}+8\pi G_{k}T_{\mu\nu}-
\Delta t_{\mu \nu}\quad,
\label{eom}
\end{eqnarray}
where the possible coordinate dependence of $G(k)$
induces an additional contribution to the stress energy tensor
\cite{Reuter:2004nx,Carroll:2004st}
\be
\Delta t_{\mu \nu}=G_{
k}\left(g_{\mu\nu}\Box-\nabla_\mu\nabla_\nu\right)\frac{1}{G_{k}}\quad.
\ee
Demanding a self consistency of those equations of motion
and a conserved stress energy tensor for matter
\be
\nabla_\mu T_{\mu \nu}=0
\ee
the following condition is found \cite{Koch:2010nn}
\be\label{condi}
R\nabla_\mu \left(\frac{1}{G(k)}\right)-
2\nabla_\mu\left(\frac{\Lambda(k)}{G(k)}\right)=0\quad.
\ee
Given a certain form of the scale dependent couplings $G(k)$
and $\Lambda(k)$, for example from the ERG approach, the above
relation allows to relate this scale $k$ to the scalar curvature $R$ of a
supposed solution.

%%%%%%%%%%%%%%%%%%%%%%%%%%%%%
\subsection{A further solution without cosmological term}
It it interesting to look for further solutions
where $g_{tt}\neq- g^{rr}$. Just adding an other
unknown function to (\ref{ansatz}) would however
leave more functions than independent equations.
Thus, it is straight forward to study scenarios
with $\Lambda(r)=0$ and
\be\label{ansatz2}
ds^2=-f(r) dt^2+
1/h(r)dr^2+
r^2 d\theta^2
+r^2 \sin (\theta)d\phi^2\quad.
\ee
With this ansatz we did not succeed to find
general solutions with variable $G(r)$.
A special solution was however
found
\bea \label{solution2}
f(r)&=&1\\ \nonumber
h(r)&=&1+\frac{2}{r c_1}\\ \nonumber
G(r)&=&c_2 \sqrt{\frac{r}{r c_1+2}}
\eea
It is quite interesting that for this solution
the adimensional coupling is the square root
of the adimensional coupling of the previous solution.

%%%%%%%%%%%%%%%%%%%%%%%%%%%%%
\subsection{Induced flow for a classical-like parameters}

\begin{figure}[hbt]
   \centering
%%\centerline{\protect\vbox{\epsfig{file=fvsr_d2_tt001_Mf1.eps,
%%width=0.6\textwidth}}}
\includegraphics[width=10cm]{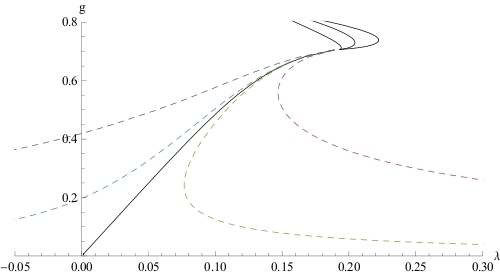}
  \caption{\label{figflowUV3} Flow of the scale dependent
  couplings $\lambda_{U}(r)$ and $g_{U}(r)$ for $g^*_U=0.707$.
  $\lambda^*_U=0.193$.
  The lower curves are plotted for $g_I=2.5$, and
  $G_0=\Sigma=1$ whith
  $l_I=\{ -0.05,\,-0.005,\,0,\,0.005,\,0.05\}$ which are depicted from left
  to right in $\{$blue,~green,~black,~orange,~red$\}$.
  The black upper curves are the real values of the classical-like scenario
  for $G_0=1$ and from left to right $M_0=\{0.7,\,1,\,2\}$.
  }
\end{figure}

Given the schematic behavior of the induced coupling flow in figure \ref{figflowUV},
one wonders how the classical-like choice of parameters fits into this picture.
The relations (\ref{c4new}, \ref{c1new}, \ref{c3new}, and \ref{c2new})
in combination with (\ref{replFP}) allow to express the dimensionless
parameters $l_I$, $\lambda_U$, $g_i$, and $g_U$
in terms of the physical parameters $G_0$, $\Lambda_0$, $M_0$ and the
undetermined scale $\Sigma$ that was introduced in order to define the dimensionless
parameters. Assuming a small physical cosmological constant $\Lambda_0$
adimensional constants that correspond to the solution (\ref{fvonrnew}-\ref{Lvonrnew}) read
approximately
\bea\label{lIclas}
l_{I,s}&\approx &-8 \sqrt{2/3}\frac{\Lambda_0}{3 \Sigma^2}+{\mathcal{O} (\Lambda_0^{4/3})}\\ \label{luclas}
\lambda_{U,s}^*&=&-4 (4/3)^{1/3}\frac{(G_0 M_0)^{5/3} \Lambda_0^{4/3}}{M}\\ \label{gIclas}
g_{I,s}&=&-\frac{1}{2}(9/2)^{1/3}\frac{G_0 \Sigma^3}{(G_0 M_0 \Lambda_0^2)^{1/3}}\\ \label{guclas}
g_{U,s}^*&=& G_0 \Sigma^2
\eea
Those adimensional constants can be combined in various ways in order to
study the fixed point behavior when varying $\Sigma$ and $M_0$.
But it allows also to form combinations which only depend on the
general physical parameters $\Lambda_0$ and $G_0$, like
$
g_{U,s}^* l_{I,s}\approx -\frac{8}{3}\sqrt{\frac{2}{3}}\Lambda_0 G_0,
$
which establishes a global relation between the UV fixed point of $g_{U,s}^*$ and the infrared
parameter $l_{I,s}$. The positivity of the gravitational coupling, further suggests
a negative value for the infrared parameter $l_I$ for positive $\Lambda_0$.
One also can ask whether the classical-like choice of parameters 
can be made compatible with the
values of the UV fixed points $\lambda^*_U=0.193$, $g^*_U=0.707$
known from the ERG approach \cite{Groh:2010ta}.
Imposing those fixed point values on the classical-like parameters (\ref{luclas}-\ref{guclas})
allows to fix the parameters $\Sigma$ and $\Lambda_0$ and leaves $G_0$ and $M_0$
as only free physical parameters. In figure \ref{figflowUV3} it is shown how the corresponding
flow would look like for the classical-like scenario.
This procedure, however gives complex values
for $\lambda(0<r<r_H)$, which strongly suggests that the classical-like scenario
is not compatible with the above fixed point values.

Please note  that (\ref{lIclas}-\ref{guclas}) is not the only possible
choice for the parameters of this solution. It only means that this choice of parameters
reproduces the standard form of the black hole metric (\ref{fstandard})
despite of the fact that $G(r)$ and $\Lambda(r)$ are not constants.
This result is new and unexpected, because it means that
from solely observing the classical de Sitter black hole metric to high precision
in the range between the horizons, one can not
conclude that $G$ and $\Lambda$ are actually constants.

%%%%%%%%%%%%%%%%%%%%%%%%%%%%%
\subsection{Horizons and Temperature}
%?? Ben ??

The horizons for this solution are given
by the conditions $f(r_H)=0$ and the black hole temperature
is obtained by using the radial derivative at this point.
This analysis reproduces only in very special cases such as the classical-like
scenario all the features that are known for the classical solution (\ref{fstandard}).
Depending on the choice of the four parameters
the horizon structure can be largely different.
In most other cases the finding of the horizons boils down to solving non-analytic equations
which only can be done numerically. This numerical study of the possible
horizons and corresponding thermodynamical behavior is postponed to a future study.

\end{appendix}
%%%%%%%%%%%%%%%%%%%%%%%%%%%%%%%%%%%%%%%%%%%%%%%%%%%%%55

%\end{thebibliography}

%\cite{Carroll:2004st}

\end{document}